\documentclass{aastex631}

\shorttitle{The flickering of RS Oph}
\shortauthors{Munari \& Tabacco}
\begin{document}

\title{Flickering returns as RS Oph reestablishes quiescent conditions following its 2021 nova outburst}

\author[0000-0001-6805-9664]{Ulisse Munari}
\affiliation{INAF Astronomical Observatory of Padova, 36012 Asiago (VI), Italy}

\author{Fulvio Tabacco}
\affiliation{ANS Collaboration, c/o Astronomical Observatory, Asiago (VI), Italy}



\begin{abstract}

RS Oph has persistently displayed flickering at optical wavelengths when
observed away from its repeating nova outbursts.  During the 2006 eruption
the flickering disappeared, and this repeated during the recent 2021 event. 
We have been monitoring RS Oph looking for the reappearance of flickering at
$B$-band following the 2021 outburst.  The flickering was still absent
($\sigma$($B$)$<$0.002 mag) on day +210 (counted from nova optical maximum),
appeared at $\sigma$($B$)=0.008 mag on day +224, and raised to
$\sigma$($B$)=0.029 mag on day +250.  On following dates the amplitude
remained large, although fluctuating.  The recovery of $B$-band quiescence
brightness by RS Oph begun around day +225 and was completed by day +260. 
The parallel patterns followed by the rise in system brightness and the
reappearance of flickering confirm the central role played in RS Oph by the
return to pre-outburst conditions of the accretion disk and the refilling by
the RG wind of the immediate circumstellar space.
 
\end{abstract}

\keywords{Recurrent Novae (1366) --- Symbiotic stars (1674) --- Stellar accretion disks (1579) --- Time series analysis (1916)}


\section{Introduction} \label{sec:intro}

RS Oph is one of the most studied symbiotic stars and recurrent novae,
following its repeated outbursts in 1898, 1933, 1958, 1967, 1985, 2006, and
2021, with possibly two more in 1907 and 1945 \citep{2010ApJS..187..275S}. 
Away from the fast evolving nova eruptions, in quiescence RS Oph is powered
by accretion onto its massive white dwarf (WD) of material lost the
red-giant companion (RG), on a 456-days orbit. 
RS Oph has been a popular target in searches for accretion-induced, rapid
light-variability (flickering) during its quiescence periods, with all
observing campaigns that have invariably detected its presence
\citep{1977MNRAS.179..587W, 1992A&A...266..237B, 1996AJ....111..414D,
2001MNRAS.326..553S, 2006AcA....56...97G, 2011BlgAJ..17...59Z,
2015MNRAS.450.3958Z, 2018MNRAS.480.1363Z, 2019BlgAJ..30...83G,
2020BlgAJ..32...35G, 2020BlgAJ..33....3G, 2021BlgAJ..34...55G}.

Flickering is not expected from RS Oph during outbursts, because of the
overwhelming glare of the eruption, and the disturbing and disrupting effect
that the impacting ejecta have on the accretion disk.  In fact, early
searches for flickering during the 2006 and 2021 outbursts detected none
\citep{2006ATel..832....1Z, 2021ATel14974....1Z, 2022ATel15296....1M}. 
After dispersal of the ejecta into the circumbinary space and the ending of
nuclear burning on the surface of the WD, the end of each outburst leaves RS
Oph significantly fainter than quiescence.  This also repeated for the 2021
outburst, as the $B$-band light-curve in Figure~1 clearly illustrates: the
drop below quiescence results from the combined effect of the cavity emptied
by the expanding ejecta which has not yet been (fully) replenished by the
wind of the RG, and of the accretion disk that has not yet returned to its
pre-outburst conditions.

As the accretion gradually resumes and the disk consequently grows in
brightness, it is expected that flickering progressively returns.  We report
here about our intensive monitoring of RS Oph in search of the return of
flickering, following the exit from Solar conjunction of the star on Jan 13,
2022 (day +156, \citet{2022ATel15169....1M}; all dates in this paper are
counted from nova optical maximum on 2021 Aug 09.58).  Some preliminar
results on the reappearance of flickering in RS Oph have been presented by
\citet{2022ATel15339....1R} and \citet{2022ATel15330....1Z}.

\begin{figure}
\includegraphics[width=18cm]{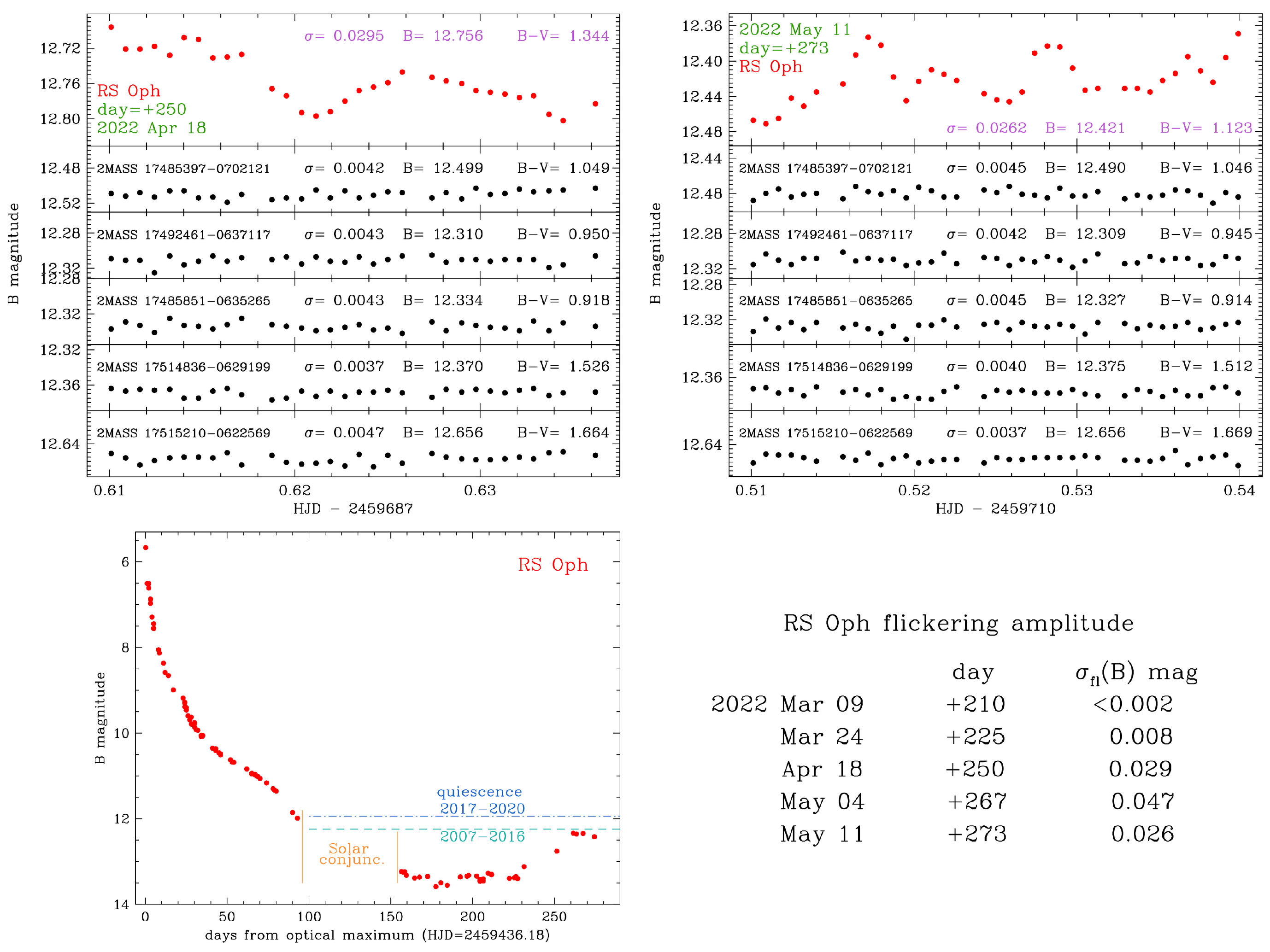}
%\plotone{Figure.pdf}
\caption{{\it Upper panels}: flickering of RS Oph and the five
check stars for two of the observing dates.  The ordinate scale is the same
in all panels.  The values for $B$, $B$$-$$V$, and $\sigma$ are in
magnitudes.  {\it Lower-left panel}: ANS Collaboration $B$-band light-curve
of RS Oph during the 2021 nova outburst.  Two levels of $B$-band brightness
in quiescence are indicated (median values from the ANS Collaboration long-term
monitoring): $B$=12.24 mag characterizing the first 10 years following the
2006 eruption, and $B$=11.94 mag for the last 5 years preceding the 2021 outburst.
{\it Lower-right panel}: summary of the results of our observations in search
for flickering in RS Oph following its emergence from Solar conjunction.
\label{fig:fig1}}
\end{figure}

\section{Observations} \label{sec:obs}

To monitor RS Oph for flickering we have used the 67/92cm Schmidt telescope
operated in Asiago by INAF National Institute of Astrophysics (Italy).  The
telescope observes robotically from Mt.  Ekar (elevation 1400m), has an
excellent optical quality and frequently enjoys 1 arcsec seeing
conditions.  The 4k$\times$4k CCD camera covers 1 squared-degree on the sky
at 0.88 arcsec/pix scale.

The observing procedure we have adopted for RS Oph is identical to that
followed by \citet{2021MNRAS.505.6121M} in their characterization of 33 new
symbiotic stars.  In essence, the observations has been conducted as
sequences of 10 exposures of 30-sec each in the $B$-filter preceded and
followed by a single 20-sec exposure with the $V$-filter, repeating the
cycle for up to 70 minutes.  Each $B$-band exposure is paired with the
closest-in-time $V$-band frame, and the transformation color equations to
the Landolt (2009) standard system are solved for such a pair against a set
of 72 reference stars imaged together with RS Oph.  These reference stars
have been selected from APASS DR8 \citep{2014CoSka..43..518H}, are
non-variable, span an interval of a couple of magnitudes centered on the
quiescence value of RS Oph, and are well distributed in color over the range
$B$$-$$V$=+0.671 to +2.048, thus well bracketing the color of RS Oph.  Around
RS Oph we have also selected a set of five field stars (not part of the
reference stars) of $B$-band brightness and $B$$-$$V$ color similar to RS
Oph, and equally well clear of interfering nearby stars.  These check stars
are measured along RS Oph with identical aperture parameters.

An example of the collected measurements is presented in Figure~1 (upper
panels).  The dispersion $\sigma$ for the five check stars (4.2 millimag on
average) represents the overall noise in the measurements (combining all
sources, from shot-noise to transformation to the standard Landolt system). 
A larger dispersion of the measurements for RS Oph ($\sigma_{RS Oph}$) is
taken as indicative of the presence of flickering, with its effective
amplitude computed as $\sigma_{fl}$=$\sqrt{\sigma^{2}_{RS Oph} -
\sigma^{2}_{CS}}$, where $\sigma_{CS}$ is the quadratic mean of $\sigma$ for
the five comparison stars.

\section{Results} \label{sec:results}

The results of our monitoring are listed at the lower-right of Figure~1,
with 2022 March 09 being the latest of our observing dates in which RS Oph
did not yet present detectable flickering.  The reappearance of flickering
coincides with the upturn of the $B$-band lightcurve around day +225. 
From day +250 the recorded amplitude of the flickering seems to depend more
on the actual observing date rather than on the $B$-band brightness of RS Oph.  

By day +260 RS Oph has resumed the same $B$-band brightness that has
characterized the initial 10 years of the quiescence following the 2006
outburst, indicating that the the immediate circumstellar space in RS OPh
has been refilled by the RG wind and the accretion disk has returned to its
pre-outburst conditions.  Such disk is probably less massive and extended
than it has been for the last five years preceding the 2021 eruption, when
the quiescence brightness of RS Oph raised by $\Delta B$=0.3 mag.  This fact
suggests that, similarly to what is observed in T CrB
\citep{2016NewA...47....7M, 2020ApJ...902L..14L}, also in RS Oph the
accretion onto the WD is not smooth and constant between outbursts, but is
instead of a more episodic nature.


%\bibliography{paper.bib}{}
\bibliography{paper.bib}{}

\begin{thebibliography}{}
\expandafter\ifx\csname natexlab\endcsname\relax\def\natexlab#1{#1}\fi
\providecommand{\url}[1]{\href{#1}{#1}}
\providecommand{\dodoi}[1]{doi:~\href{http://doi.org/#1}{\nolinkurl{#1}}}
\providecommand{\doeprint}[1]{\href{http://ascl.net/#1}{\nolinkurl{http://ascl.net/#1}}}
\providecommand{\doarXiv}[1]{\href{https://arxiv.org/abs/#1}{\nolinkurl{https://arxiv.org/abs/#1}}}

\bibitem[{{Bruch}(1992)}]{1992A&A...266..237B}
{Bruch}, A. 1992, \aap, 266, 237

\bibitem[{{Dobrzycka} {et~al.}(1996){Dobrzycka}, {Kenyon}, \&
  {Milone}}]{1996AJ....111..414D}
{Dobrzycka}, D., {Kenyon}, S.~J., \& {Milone}, A. A.~E. 1996, \aj, 111, 414,
  \dodoi{10.1086/117794}

\bibitem[{{Georgiev} {et~al.}(2019){Georgiev}, {Zamanov}, {Boeva}, {Latev},
  {Spassov}, {Mart{\'\i}}, {Nikolov}, {Ibryamov}, {Tsvetkova}, \&
  {Stoyanov}}]{2019BlgAJ..30...83G}
{Georgiev}, T.~B., {Zamanov}, R.~K., {Boeva}, S., {et~al.} 2019, Bulgarian
  Astronomical Journal, 30, 83

\bibitem[{{Georgiev} {et~al.}(2020{\natexlab{a}}){Georgiev}, {Zamanov},
  {Boeva}, {Latev}, {Spassov}, {Mart{\'\i}}, {Nikolov}, {Ibryamov},
  {Tsvetkova}, \& {Stoyanov}}]{2020BlgAJ..32...35G}
---. 2020{\natexlab{a}}, Bulgarian Astronomical Journal, 32, 35

\bibitem[{{Georgiev} {et~al.}(2020{\natexlab{b}}){Georgiev}, {Zamanov},
  {Boeva}, {Latev}, {Spassov}, {Mart{\'\i}}, {Nikolov}, {Ibryamov},
  {Tsvetkova}, \& {Stoyanov}}]{2020BlgAJ..33....3G}
---. 2020{\natexlab{b}}, Bulgarian Astronomical Journal, 33, 3

\bibitem[{{Georgiev} {et~al.}(2021){Georgiev}, {Zamanov}, {Boeva}, {Latev},
  {Spassov}, {Mart{\'\i}}, {Nikolov}, {Ibryamov}, {Tsvetkova}, \&
  {Stoyanov}}]{2021BlgAJ..34...55G}
---. 2021, Bulgarian Astronomical Journal, 34, 55

\bibitem[{{Gromadzki} {et~al.}(2006){Gromadzki}, {Mikolajewski}, {Tomov},
  {Bellas-Velidis}, {Dapergolas}, \& {Galan}}]{2006AcA....56...97G}
{Gromadzki}, M., {Mikolajewski}, M., {Tomov}, T., {et~al.} 2006, \actaa, 56, 97

\bibitem[{{Henden} \& {Munari}(2014)}]{2014CoSka..43..518H}
{Henden}, A., \& {Munari}, U. 2014, Contributions of the Astronomical
  Observatory Skalnate Pleso, 43, 518

\bibitem[{{Luna} {et~al.}(2020){Luna}, {Sokoloski}, {Mukai}, \& {M.
  Kuin}}]{2020ApJ...902L..14L}
{Luna}, G. J.~M., {Sokoloski}, J.~L., {Mukai}, K., \& {M. Kuin}, N.~P. 2020,
  \apjl, 902, L14, \dodoi{10.3847/2041-8213/abbb2c}

\bibitem[{{Marchev} {et~al.}(2022){Marchev}, {Pavlova}, \&
  {Zamanov}}]{2022ATel15296....1M}
{Marchev}, D., {Pavlova}, N., \& {Zamanov}, R. 2022, The Astronomer's Telegram,
  15296, 1

\bibitem[{{Munari} {et~al.}(2016){Munari}, {Dallaporta}, \&
  {Cherini}}]{2016NewA...47....7M}
{Munari}, U., {Dallaporta}, S., \& {Cherini}, G. 2016, \na, 47, 7,
  \dodoi{10.1016/j.newast.2016.01.002}

\bibitem[{{Munari} {et~al.}(2022){Munari}, {Valisa}, \&
  {Dallaporta}}]{2022ATel15169....1M}
{Munari}, U., {Valisa}, P., \& {Dallaporta}, S. 2022, The Astronomer's
  Telegram, 15169, 1

\bibitem[{{Munari} {et~al.}(2021){Munari}, {Traven}, {Masetti}, {Valisa},
  {Righetti}, {Hambsch}, {Frigo}, {{\v{C}}otar}, {De Silva}, {Freeman},
  {Lewis}, {Martell}, {Sharma}, {Simpson}, {Ting}, {Wittenmyer}, \&
  {Zucker}}]{2021MNRAS.505.6121M}
{Munari}, U., {Traven}, G., {Masetti}, N., {et~al.} 2021, \mnras, 505, 6121,
  \dodoi{10.1093/mnras/stab1620}

\bibitem[{{Romanov}(2022)}]{2022ATel15339....1R}
{Romanov}, F. 2022, The Astronomer's Telegram, 15339, 1

\bibitem[{{Schaefer}(2010)}]{2010ApJS..187..275S}
{Schaefer}, B.~E. 2010, \apjs, 187, 275, \dodoi{10.1088/0067-0049/187/2/275}

\bibitem[{{Sokoloski} {et~al.}(2001){Sokoloski}, {Bildsten}, \&
  {Ho}}]{2001MNRAS.326..553S}
{Sokoloski}, J.~L., {Bildsten}, L., \& {Ho}, W. C.~G. 2001, \mnras, 326, 553,
  \dodoi{10.1046/j.1365-8711.2001.04582.x}

\bibitem[{{Walker}(1977)}]{1977MNRAS.179..587W}
{Walker}, A.~R. 1977, \mnras, 179, 587, \dodoi{10.1093/mnras/179.4.587}

\bibitem[{{Zamanov}(2011)}]{2011BlgAJ..17...59Z}
{Zamanov}, R. 2011, Bulgarian Astronomical Journal, 17, 59

\bibitem[{{Zamanov} {et~al.}(2022){Zamanov}, {Marchev}, {Marchev}, {Atanasova},
  \& {Pavlova}}]{2022ATel15330....1Z}
{Zamanov}, R., {Marchev}, V., {Marchev}, D., {Atanasova}, T., \& {Pavlova}, N.
  2022, The Astronomer's Telegram, 15330, 1

\bibitem[{{Zamanov} {et~al.}(2006){Zamanov}, {Panov}, {Boer}, \& {Le
  Coroller}}]{2006ATel..832....1Z}
{Zamanov}, R., {Panov}, K., {Boer}, M., \& {Le Coroller}, H. 2006, The
  Astronomer's Telegram, 832, 1

\bibitem[{{Zamanov} {et~al.}(2021){Zamanov}, {Stoyanov}, {Kostov}, {Boeva},
  {Moyseev}, {Marti}, \& {Luque-Escamilla}}]{2021ATel14974....1Z}
{Zamanov}, R., {Stoyanov}, K., {Kostov}, A., {et~al.} 2021, The Astronomer's
  Telegram, 14974, 1

\bibitem[{{Zamanov} {et~al.}(2015){Zamanov}, {Latev}, {Boeva}, {Sokoloski},
  {Stoyanov}, {Bachev}, {Spassov}, {Nikolov}, {Golev}, \&
  {Ibryamov}}]{2015MNRAS.450.3958Z}
{Zamanov}, R., {Latev}, G., {Boeva}, S., {et~al.} 2015, \mnras, 450, 3958,
  \dodoi{10.1093/mnras/stv873}

\bibitem[{{Zamanov} {et~al.}(2018){Zamanov}, {Boeva}, {Latev}, {Mart{\'\i}},
  {Boneva}, {Spassov}, {Nikolov}, {Bode}, {Tsvetkova}, \&
  {Stoyanov}}]{2018MNRAS.480.1363Z}
{Zamanov}, R.~K., {Boeva}, S., {Latev}, G.~Y., {et~al.} 2018, \mnras, 480,
  1363, \dodoi{10.1093/mnras/sty1816}

\end{thebibliography}
\bibliographystyle{aasjournal}



\end{document}